\documentclass{article}
\usepackage{spconf,amsmath,amssymb,graphicx,hyperref,booktabs,cite}
\usepackage{xcolor}
\usepackage{makecell}
\usepackage{enumitem}
\usepackage{tipa}
\usepackage{subfig}

\usepackage{bbding}
\usepackage{fancyhdr}
\fancypagestyle{firstpage}{
\fancyhf{}
\fancyfoot[C]{\footnotesize Copyright 2026 IEEE. Published in ICASSP 2026 - 2026 IEEE International Conference on Acoustics, Speech and Signal Processing (ICASSP), scheduled for 3-8 May 2026 in Barcelona, Spain. Personal use of this material is permitted. However, permission to reprint/republish this material for advertising or promotional purposes or for creating new collective works for resale or redistribution to servers or lists, or to reuse any copyrighted component of this work in other works, must be obtained from the IEEE. Contact: Manager, Copyrights and Permissions / IEEE Service Center / 445 Hoes Lane / P.O. Box 1331 / Piscataway, NJ 08855-1331, USA. Telephone: + Intl. 908-562-3966.}

}
\title{Interpretable Modeling of Articulatory Temporal Dynamics from real-time MRI for Phoneme Recognition}
\name{Jay Park$^{1*}$, 
Hong Nguyen$^{1*}$, 
Sean Foley$^{1,2}$, 
Jihwan Lee$^1$, 
\\\em{Yoonjeong Lee$^1$, 
Dani Byrd$^2$, 
Shrikanth Narayanan$^{1,2}$}
}

\address{$^1$Signal Analysis and Interpretation Lab, University of Southern California \\
$^2$Department of Linguistics, University of Southern California 
}

\begin{document}
\maketitle
\begin{abstract}
Real-time Magnetic Resonance Imaging ({rtMRI}) visualizes vocal tract action, offering a comprehensive window into speech articulation. However, its signals are high dimensional and noisy, hindering interpretation. We investigate compact representations of spatiotemporal articulatory dynamics for phoneme recognition from midsagittal vocal tract rtMRI videos. We compare three feature types: (1) raw video, (2) optical flow, and (3) six linguistically-relevant regions of interest (ROIs) for articulator movements. We evaluate models trained independently on each representation, as well as multi-feature combinations. Results show that multi-feature models consistently outperform single-feature baselines, with the lowest phoneme error rate (PER) of 0.34 obtained by combining ROI and raw video. Temporal fidelity experiments demonstrate a reliance on fine-grained articulatory dynamics, while ROI ablation studies reveal strong contributions from tongue and lips. Our findings highlight how rtMRI-derived features provide accuracy and interpretability, and establish strategies for leveraging articulatory data in speech processing. The source code is publicly available.\footnote{\url{https://github.com/usc-sail/multimodal_av_temporal_phoneme}}
\end{abstract}

\section{Introduction}\label{sec:intro}

\begin{figure*}[t!]
\centering
  \includegraphics[width=0.9\linewidth]{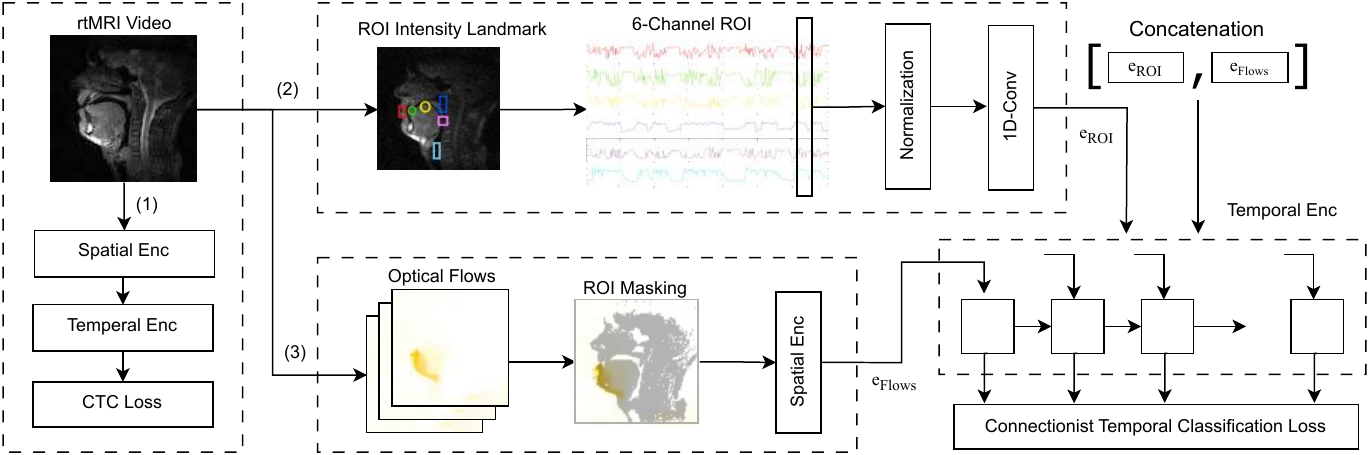}
  \caption{Overview of three model pipelines for analyzing spatial–temporal articulatory dynamics in rtMRI. (1) Raw video frames, (2) Six-channel ROI pixel intensity, and (3) Optical flow encodes fine-grained articulatory kinematics.}
  \label{fig:pipelines}
\end{figure*}

Speech production is fundamentally articulatory; phonemes emerge from the coordination of constriction actions in the vocal tract. These actions, defined by constriction location and degree, involve articulators such as the lips, tongue, velum, and larynx \cite{browman1992articulatory}. Neural activity in the human sensorimotor cortex is highly correlated with these articulatory gestures \cite{CHARTIER20181042}, underscoring their central role in both cognitive and motor models of speech. 

Capturing articulatory dynamics provides not only a scientific lens on human speech but also a powerful complement to acoustic-based processing. Articulatory data have been obtained with instruments such as Electromagnetic Articulography (EMA) and ultrasound; however these capture only subsets of vocal tract variables —e.g., EMA lacks data tracking from the velum, tongue root, and larynx. Real-time magnetic resonance imaging (rtMRI) provides a fuller view of the vocal tract, including the posterior tongue root and laryngeal regions \cite{Narayanan2004Anapproachtoreal-time,Lingala2016Afastandflexible}. 
Midsagittal rtMRI has enabled applications such as vowel-consonant-vowel and phoneme sequence prediction \cite{saha2018automaticspeechidentificationvocal, foley2025disentanglingcontributionsarticulationacoustics}, speech-to-rtMRI synthesis \cite{Nguyen2024Speech2rtMRISD}, and rtMRI-to-speech synthesis \cite{Shah2024MRI2SpeechSS}. However, unprocessed rtMRI video frames, while spatially rich, lack structured interpretable articulatory features and are high dimensional and noisy, complicating both training and interpretation.

While some studies have fed raw rtMRI frames directly into speech processing models in various applications, this approach suffers from high dimensionality and limited interpretability. Other articulator tracking modalities such as EMA, which has excellent temporal resolution but relies on limited point-tracking, have emphasized articulatory dynamics more explicitly and that success highlights the value of focusing on dynamics rather than static frames \cite{Toutios2011EstimatingTC, Toutios2013ArticulatorySO, BYRD1998173}. Recent work has shown a strong correlation between self-supervised acoustic representations and vocal tract dynamics \cite{10094711}, underscoring the centrality of articulatory motion in speech modeling. However, EMA provides only sparse sampling of the vocal tract and misses tongue root, velum and laryngeal action \cite{Toutios2016AdvancesIR}. rtMRI, on the other hand, captures a full mid-sagittal view of the vocal tract but demands methods that can distill its high-dimensional frames into compact, interpretable representations.

Given these limitations, our study proposes effective preprocessing strategies for embedding articulatory motion in rtMRI videos for downstream tasks such as phoneme sequence prediction. We address two research questions: (1) How can speech-relevant information best be extracted from rtMRI video while balancing dimensionality and informativity? and (2) Which articulatory features are most critical for phoneme sequence prediction? To address these, we evaluate preprocessing methods that encode articulatory motion into low-dimensional yet informative representations. Such representations can both improve recognition accuracy and enhance model interpretability, while reducing training complexity and risk of overfitting.
Our contributions are:
\begin{itemize}[left=0pt, topsep=0pt, partopsep=0pt, parsep=0pt, itemsep=1pt]
    \item We evaluate optical flow and ROI intensity as representations of spatial and temporal information in rtMRI videos. While raw video alone yields the lowest phoneme error rate (PER) among single-feature models, combining feature types consistently outperforms single inputs.
    \item We show that phoneme sequence prediction from six-channel ROI features depends most strongly on lips, tongue tip and tongue root. Temporal fidelity tests of rtMRI features through shuffle, reversal, and resampling further demonstrate that accurate phoneme recognition relies on specific fine-grained articulatory dynamics.
\end{itemize}

Together, these findings demonstrate how targeted rtMRI preprocessing yields compact yet informative articulatory representations, advancing both interpretable speech modeling and efficient phoneme recognition.






\vspace{-8pt}
\section{Methods}
We examine spatial-temporal articulatory dynamics using two different methods: ROI pixel intensity maps and optical flow. Let the articulatory state at time $t$ be $s_t \in \mathcal{S} \subset \mathbb{R}^{W \times H}
$ (rtMRI domain). Observations are modality-specific mappings $h_m (s_t)$ in which $h_\text{flows}$ and $h_\text{ROI}$ are defined as optical flow representation and ROI pixel intensity, respectively. We hypothesize that $h_\text{flows}$  captures high spatial resolution and directly encodes kinematics in rtMRI videos while $h_\text{ROI}$ acts as a pooled, low-rank approximation of articulator dynamics robust to noise (averaging) but lacking fine-grained geometry. Finally, we provide experiments to verify these hypotheses.
\vspace{-5pt}
\subsection{Spatial-temporal dynamics via ROI Pixel Intensity}
\label{ssec:subhead}

Using the VocalTract ROI Toolbox in MATLAB \cite{blaylock2021vocaltract}, we define six channels of regions of interest (ROIs) $\{\mathcal{R}_c\}_{c=1}^{6} \subset \Omega$: lip aperture (LA), tongue tip (TT), tongue body (TB), velum (VL), tongue root (TR), and larynx (LX). Regions were hand-selected by trained speech scientists based on pertinent vocal tract kinematics and spatial landmarks. Per-channel normalized intensity, as shown in Figure ~\ref{fig:pipelines}, is derived as $y_{t,c}^\text{ROI} = \frac{1}{|\mathcal{R}_c|} \int_{\mathcal{R}_c} \mathcal{I}_t (x) dx$. Therefore, the observation operator is a stacked normalized vector $h_\text{ROI}(s_t) = y_{t}^\text{ROI} = [y_{t,1}, y_{t,2},.. y_{t,c}] \in \mathbb{R}^{t \times c}$, 
where $t$ and $c$ represent the temporal and spatial dimensions of the articulators, respectively. These features are processed with two layers of 1D convolution and a temporal encoder (LSTM/Mamba \cite{gu2024mambalineartimesequencemodeling}) to generate articulatory embeddings. Finally, these embeddings are passed through a lightweight classification head to generate phoneme probability distributions (Figure ~\ref{fig:pipelines}).
\vspace{-15pt}
\subsection{Spatial-temporal dynamics via Optical Flow}
Optical flow \cite{dong2024memflowopticalflowestimation} is an alternative representation for modeling temporal dynamics in video sequences. Traditional optical flow methods \cite{Lucas1981AnII} capture pixel-level motion but produce noisy representations at the object level, particularly in rtMRI video. While no deep learning models have been trained on medical images due to the lack of ground-truth annotations, our qualitative evaluation suggests that MemFlow \cite{dong2024memflowopticalflowestimation}, whose checkpoint was pretrained on the Sintel dataset \cite{10.1007/978-3-642-33783-3_44}, captures object-level motion patterns in rtMRI to some extent. Therefore, we extract optical flow fields $\mathcal{F}_t(x) \in \mathbb{R}^{T,C,W,H}$ from video set
$v$, where C = 2 represents horizontal and vertical motion components. To stabilize noise in the estimated optical flow, we suppress irrelevant background motion by applying a mask derived from intensity thresholding of the original images, thereby retaining only articulator-related movements. The masked flows are subsequently encoded with a Vision Transformer \cite{10.5555/3295222.3295349} to obtain spatial flow embeddings $e_\text{flows} \in \mathbb{R}^{T, D}$. Temporal dependencies are then modeled using a temporal encoder (Mamba/LSTM), and the output embeddings are passed through a lightweight classification head to generate phoneme probability distributions in a way identical to the ROI approach (Figure ~\ref{fig:pipelines}).

\label{sec:format}

\begin{table}
  \caption{PER of comparison models from \cite{foley2025disentanglingcontributionsarticulationacoustics} (top), PER from uni-feature models in current study (middle), and PER from multi-feature models in current study (bottom). The best scores for each metric are in bold. DIV: Training Divergence, N/A: Not reported, \Snowflake : Frozen weights.}
  \label{percomparison}
  \centering
  \resizebox{\columnwidth}{!}{%
  \begin{tabular}{l c c c c}
    \toprule
    \textbf{\makecell{rtMRI\\Features}} & \textbf{\makecell{Spatial \\Enc}} & \textbf{\makecell{Temp \\Enc}} & \textbf{PER}$\downarrow$ & \makecell{\textbf{Phone Acc}$\uparrow$ \\ Top-1 \hspace{1pt} Top-3}\\
    \midrule
    \multicolumn{4}{l}{\textit{Results reported by \cite{foley2025disentanglingcontributionsarticulationacoustics}}} \\
    Raw Video & ViT-B \Snowflake &\makecell{LSTM + \\ Conformer} & 0.49 & N/A\\
    \midrule
    \multicolumn{4}{l}{\textit{Uni-Feature}} \\
    \text{Raw Video} & ViT-B & Mamba & DIV & -\\
    & ResNet & Mamba & 0.37 & 11.7 \hspace{1pt} 67.6 \\
    & ResNet & LSTM & 0.38 & 12.8 \hspace{1pt} \textbf{67.7}\\
    \text{Flows} & ResNet & Mamba & \text{0.41} & 11.2 \hspace{1pt} 67.5  \\
    & ResNet & LSTM & 0.43 & \hspace{1pt} 8.6 \hspace{2pt} 44.4\\
    \text{6C-ROI} & Conv1D & LSTM & \text{0.53} & 12.2 \hspace{1pt} 49.9\\
    & Conv1D & Mamba & 0.61 & \hspace{1pt} 9.2 \hspace{2pt} 48.9\\
    \midrule
    \multicolumn{3}{l}{\textit{Multi-Feature}} \\
    \multicolumn{3}{l}{\text{6C-ROI + Flows}} & 0.35 & 15.1 \hspace{1pt} 66.5\\
    \multicolumn{3}{l}{\text{6C-ROI + Raw Video}}  & \textbf{0.34} & 14.8 \hspace{1pt} 64.1 \\
    \multicolumn{3}{l}{\text{Raw Video + Flows}}   & 0.39 & \textbf{15.3} \hspace{1pt} 66.4 \\
    \bottomrule
  \end{tabular}}
\end{table}

\vspace{-10pt}
\section{Experimental Setup}
\label{sec:pagestyle}
\vspace{-8pt}
\subsection{Dataset}
\label{ssec:subhead}
We utilize an rtMRI audio/video dataset of one male American English speaker presented in \cite{seanunpublished}. Speech in this dataset, the "longest single speaker" corpus available to date, contains the 460 phrases from the USC TIMIT \cite{Narayanan2013USCTIMITA} corpus, the phrases from the USC 75-Speaker Dataset \cite{lim2021multispeaker}, and spontaneous response to questions about daily activities. This dataset consists of approximately one hour of speech recorded using a midsagittal orientation of the vocal tract with a 0.55T MRI scanner and a custom receiver coil for the upper airway \cite{Muoz2022EvaluationOA}. The dataset consists of 71 videos with a frame rate of 99 frames/sec (FPS) \cite{Kumar2024StateoftheartSP}, audio with a sampling rate of 16 kHz, and phoneme-level annotations.
\vspace{-10pt}
\subsection{Preprocessing}
\label{ssec:subhead}
Videos were resampled to 100 FPS to simplify preprocessing. Audio and video sequences were sliced into 5 second samples with a 2.5 second sliding window, resulting in 1224 samples and a total duration of 102 minutes. 924 samples from videos 18-71 were used for training, 100 samples from videos 13-17 were used for validation, and 100 samples from videos 1-12 were used for testing. Video samples $\mathbf{V} \in \mathbb{R}^{500 \times 104 \times 104}$ were then converted to optical flow and ROI representations.

\begin{table}[t]
  \caption{PER increase from temporal fidelity tests. Features with the highest increase in PER for each test type are in bold.}
  \vspace{4pt}
  \label{temporalfidelity}
  \centering
  \begin{tabular}{l c c c c c}
    \toprule
    \textbf{\makecell{rtMRI \\ Features}} & \textbf{\makecell{Cross-\\ Phon.}} & \textbf{\makecell{Per-\\Phon.}} & \textbf{\makecell{Time-\\Rev.}} & \textbf{\makecell{Up \\ Samp.}}  & \textbf{\makecell{Down \\ Samp.}} \\
    \midrule
    \text{Raw Vid.} & 0.28 & \textbf{0.21} & \textbf{0.25} & \textbf{0.30} & 0.20\\
    \text{Flows} & \textbf{0.48} & 0.04 & 0.07 & 0.22 & \textbf{0.24}\\
    \text{6C-ROI} & 0.07 & {0.15} & 0.12 & 0.02 & 0.03\\
    \bottomrule
  \end{tabular}
\end{table}

\begin{table}[t]
  \caption{PER increase of five-channel ROI model upon removing articulatory features. Largest changes are in bold.}
  \vspace{4pt}
  \label{tab:roi_ablation}
  \centering
  \begin{tabular}{cccccc}
    \toprule
    \textbf{LA} & \textbf{TT} &\textbf{TD} & \textbf{VL} & \textbf{TR} & \textbf{LX}\\
    \midrule
      {0.13} 
     &  \textbf{0.15}  &  0.10  &  0.03  &  0.06  & 0.08 \\
    \bottomrule
  \end{tabular}
\end{table}

\subsection{Training}
\label{ssec:subhead}
For phoneme sequence prediction on running speech, we use the Connectionist Temporal Classification (CTC) loss function \cite{10.1145/1143844.1143891} to train all of our models. Utilizing an NVIDIA A40 GPU, we train with a batch size of 1 and an Adam optimizer with an initial learning rate of $10^{-3}$ and a weight decay of $10^{-4}$ to regularize parameters. The learning rate was scheduled to decay by a factor of 0.9 every 20 epochs over 300 total epochs using a step scheduler. While \cite{foley2025disentanglingcontributionsarticulationacoustics} compared PER for models trained on audio, raw video, and both combined, we measure performance of models trained independently on raw video, optical flow and vocal tract ROI, as well as for models combining pairs of these features.

\subsection{Evaluation metrics}
\label{ssec:subhead}

We evaluate model performance using two metrics: sequence-level PER and per-phoneme classification accuracy (Top-1 and Top-3) on the test set. For each phoneme $y_i$, we calculate Top-1 and Top-3 per-phoneme classification accuracy $\textbf{P}_{\text{Acc}} = \frac{\#\text{correct frames for } y_i}{\text{Total frames for } y_i}$ to show which phonemes are problematic (e.g., the model struggles with \textipa{/d/}, but not with \textipa{/s/}). Confusion matrices show the associated substitution patterns (e.g., \textipa{/p/} and \textipa{/b/}, \textipa{/s/} and \textipa{/S/}).

\section{Results and Discussion}
\label{sec:typestyle}

\subsection{Comparison of features}
\label{ssec:subhead}

PER and per-phoneme classification accuracy (Top-1 and Top-3) for each speech representation are shown in Table ~\ref{percomparison}. Since \cite{foley2025disentanglingcontributionsarticulationacoustics} predicted phoneme sequences with the same rtMRI corpus, we include their results for comparison although our testing dataset is different. Models trained independently on raw video, optical flow, and six-channel ROI have the best PER values of 0.37, 0.41, and 0.53 respectively. While Top-1 per-phoneme classification accuracy remains consistent for all features, Top-3 per-phoneme classification accuracy is best with raw video and optical flow models.

\begin{figure}\centering
\subfloat{\label{fig:maeea}\includegraphics[width=.51\linewidth]{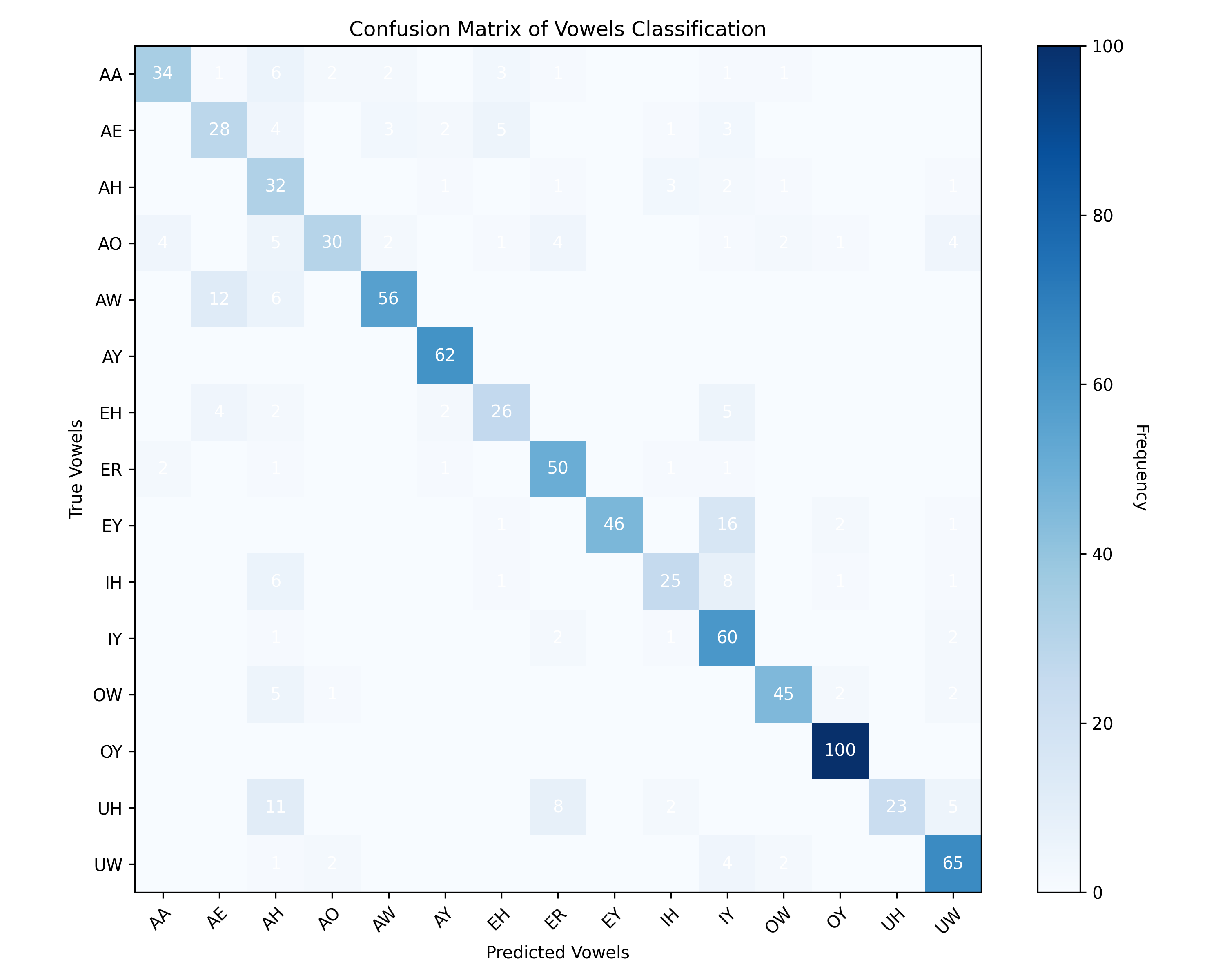}}\hfill
\subfloat{\label{fig:maeeb}\includegraphics[width=.49\linewidth]{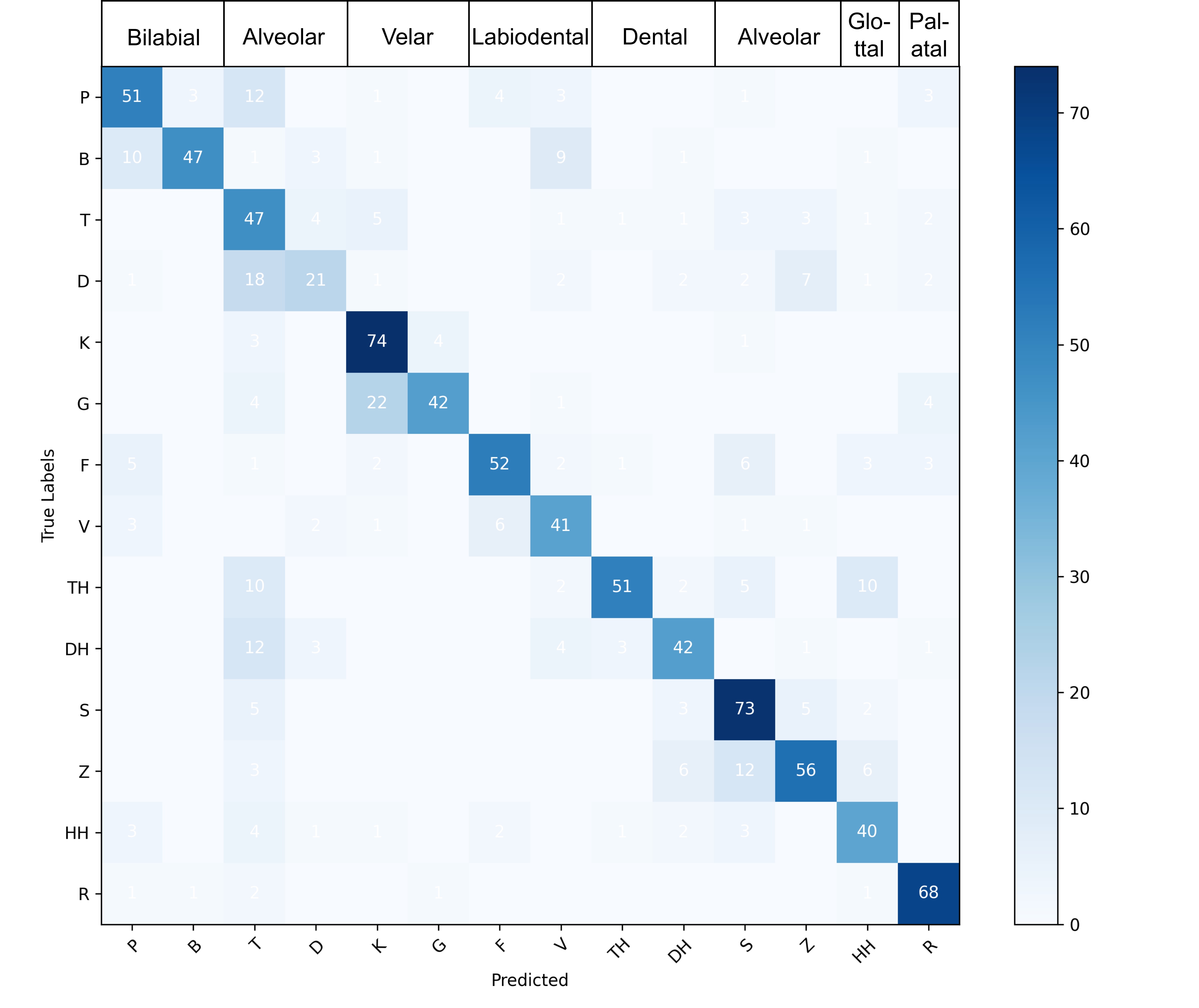}}\hfill
\caption{Confusion matrices in percentage of 6-Channel ROI + raw video showing vowels (left) and consonants (right). We ignore the Palatal Consonants /SH/ and /ZH/ since sample size for each is low.}
\label{fig:confusionmatrices}
\end{figure}

The results for PER and per-phoneme classification accuracy of multimodal models trained on pairs of rtMRI-acquired features are also presented in Table ~\ref{percomparison}. We also present confusion matrices for our best-performing model trained on six-channel ROI and raw video in Figure ~\ref{fig:confusionmatrices}. These matrices show that most phonemes are recognized to varying degrees apart from \textipa{/Z/} and \textipa{/S/}. The matrices also show errors between similar consonants (e.g., velars /k/ and /g/). While Top-3 per-phoneme classification accuracy fails to exceed that of single feature models, Top-1 per-phoneme classification accuracy is improved by using multiple features. The lowest PER of 0.34 was acquired with a multimodal model trained on six-channel ROI and raw video, indicating that these two feature types provide complementary information.

\begin{figure}[t!]
\centering
\includegraphics[width=1\linewidth]{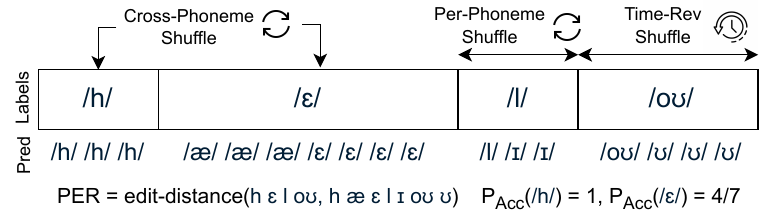}
  \caption{Shuffle tests for temporal fidelity evaluation with example of PER and P-Acc calculation.}
  \label{fig:temporalfidelity}
\end{figure}

\vspace{-5pt}
\subsection{Temporal fidelity}
We investigate the role of temporal dynamics in rtMRI by rearranging frames using five methods (Figure ~\ref{fig:temporalfidelity}):
\begin{enumerate}[left=0pt, topsep=0pt, partopsep=0pt, parsep=0pt, itemsep=1pt]
    \item \textbf{Cross-phoneme shuffle:} Phoneme order is shuffled while frame order within each phoneme is maintained. This approach tests the relevance of phoneme order.
    \item \textbf{Per-phoneme shuffle:} Frame order is shuffled within each phoneme while phoneme order is maintained to test the relevance of temporal dynamics within each phoneme.
    \item \textbf{Time-reversal:} The order of frames is reversed in sequence within each phoneme. This approach preserves smoothness while flipping direction of motion. This approach tests sensitivity to causal dynamics.
    \item \textbf{Upsampling:} Frames are upsampled by a factor of two.
    \item \textbf{Downsampling:} Frames are downsampled by a factor of two to test sensitivity to temporal resolution.
\end{enumerate}

We present results for our temporal fidelity analysis in Table ~\ref{temporalfidelity}. The increase in PER with models retrained on shuffled data for all video features suggests that embeddings rely heavily on temporal continuity and cues from coarticulation. The PER increases of 0.20, 0.24, and 0.03 upon downsampling raw video, Optical Flow, and ROI respectively demonstrate sensitivity to high frequency components. Meanwhile, the PER increases of 0.30, 0.22, and 0.02 upon upsampling raw video, Optical Flow, and ROI respectively may be attributed to model overfitting with a larger number of parameters.

\subsection{Ablation: Importance of each ROI}
\label{ssec:subhead}
We explore which articulators play a more critical role in capturing articulatory motion. Adopting a similar methodology to \cite{Lee2025ArticulatoryFP}, we train a model with each ROI removed and then calculate the PER difference compared to the full model.
The PER increases after removing each channel are shown in Table ~\ref{tab:roi_ablation}. The largest changes occur when removing TT and LA regions, with PER increases of 0.13 and 0.15 respectively. This implies that articulators are complementary, with especially strong contributions from the information associated with tongue tip constriction and bilabial action.

\section{Conclusion and Future Work}

Our study shows that rtMRI preprocessing choices strongly shape both accuracy and interpretability in phoneme recognition. Multi-feature models (e.g., ROI + raw video) consistently outperform single inputs: ROI features provide compact, interpretable summaries, while raw video and optical flow preserve richer kinematic detail. Confusion analyses highlight errors in phonetically similar categories, temporal fidelity tests confirm reliance on within-phoneme dynamics, and ablations reveal the central role of lip and tongue movements. Together, these findings underscore the complementarity of feature types, with combined representations offering both efficiency and performance.

As a last note, given that manual ROI extraction is time consuming, future work can seek to automate ROI feature extraction using technically optimal strategies. A key limitation of our work is our reliance on one subject, restricting cross-subject generalization and lessening the need for cross-subject video and region  normalization for different vocal tracts. Consequently, our models may overfit to individual articulators. The dataset also lacks certain phonemes (\textipa{/Z/} in particular), thereby increasing PER. Future studies should expand to multi-subject datasets with phoneme-aligned labels and vocal-tract normalization for constriction location and degree to improve generalizability. 

\section{Acknowledgement}

This work was supported by The U.S. National Science Foundation under Grant NSF (IIS-2311676, BCS-2240349, RI-2106930). We would like to thank Prof. Sudarsana Reddy Kadiri for his valuable comments and insights.

\bibliographystyle{IEEEbib}
\small
\bibliography{strings,refs}

\end{document}